# Scalable global grid catalogue for Run3 and beyond

M Martinez Pedreira[1] and C Grigoras[1] for the ALICE Collaboration

[1] CERN, CH-1211, Geneva 23 Switzerland

E-mail: miguel.martinez.pedreira@cern.ch, costin.grigoras@cern.ch

**Abstract.** The AliEn (ALICE Environment) file catalogue is a global unique namespace providing mapping between a UNIX-like logical name structure and the corresponding physical files distributed over 80 storage elements worldwide. Powerful search tools and hierarchical metadata information are integral parts of the system and are used by the Grid jobs as well as local users to store and access all files on the Grid storage elements. The catalogue has been in production since 2005 and over the past 11 years has grown to more than 2 billion logical file names. The backend is a set of distributed relational databases, ensuring smooth growth and fast access. Due to the anticipated fast future growth, we are looking for ways to enhance the performance and scalability by simplifying the catalogue schema while keeping the functionality intact. We investigated different backend solutions, such as distributed key value stores, as replacement for the relational database. This contribution covers the architectural changes in the system, together with the technology evaluation, benchmark results and conclusions.

## 1. Introduction

The ALICE [1] experiment at CERN is one of the four big particle detectors in the Large Hadron Collider (LHC). Its main goal is to study heavy-ion (Pb-Pb) collisions at high center of mass energies. It is expected that the temperatures and energy density lead to the formation of a new phase of matter, the quark-gluon plasma, a state of the matter where quarks and gluons are freed, conditions that are believed to have existed a fraction of second after the Big Bang. The existence of such a phase and its properties are key issues in Quantum Chromodynamics (QCD) for the understanding of confinement-deconfinement and of chiral-symmetry phase transitions.

During data taking periods there are huge amounts of data gathered, in the order of tens of petabytes each year. The quantity of information is so big, that a new computing infrastructure was created for the purpose of storing and analyzing it, since a unique big computing resource would not suffice to process it. ALICE members make use of the Worldwide LHC Computing Grid (WLCG) [2]. It unites 170 computing centers in 42 countries, providing global computing resources to store, distribute and analyze the petabytes of data coming from the LHC experiments.

ALICE Environment (AliEn) [3] is the middleware framework used on top of WLCG within ALICE, and coordinates the computing operations and data management of the collaboration in the grid. The AliEn File Catalogue is a MySQL-based database that holds the information of every single file in the ALICE grid. As of 2016, it contains more than 2 billion logical entries, and more than 3 billion physical location pointers. The catalogue growth over the time is constant and smooth, but we will face a significant increase on the amount of computing resources in the next LHC run, thus we need to adapt our system to cope with the new requirements.

For this reason, a new catalogue is being developed, making use of new technologies both from the point of view of software and hardware. Some framework changes will be introduced to optimize parts of its workflows. We present the details and performance tests in the next sections.

## 2. Parameters of the present File Catalogue

The current catalogue grew by a factor of 1.62 between July 2015 and July 2016, and 2.65 in the last 3 years. The growth has shown to be smooth and constant, with some periods where there are peaks in the file registrations, especially during Pb-Pb data taking and processing. The current number of logical files surpasses 2 billion while the physical files reach almost 3 billion.

When it comes to performance, reads come as the most demanded operation, but other operations are significant as well.

| MySQL query rates | | | |
|---|---|---|---|
| Command | Average rate | Max rate | Group rate |
| INSERT | 155 Hz | 4747 Hz | 577 Hz |
| INSERT SELECT | 115 Hz | 1711 Hz | |
| REPLACE | 79 Hz | 957 Hz | |
| UPDATE | 228 Hz | 2339 Hz | |
| SELECT | 4802 Hz | 23333 Hz | 11930 Hz |
| SELECT (cached) | 7128 Hz | 35470 Hz | |
| DELETE | 291 Hz | 6799 Hz | 291 Hz |
| TOTAL | 12798 Hz | | |

**Figure 1.** MySQL query rates January-July 2016

In Figure 1 the numbers represent the queries arriving to MySQL directly, where the group rate column contains the sum of the averages by query type. However, we make use of a cache layer, that filters a fraction of the read-only queries, to reduce the load on the server while giving faster responses. In particular, the cache deals with file information and locations (whereis), lookups for files in the hierarchy (find) and permission requests (access). Some of these operations can be very heavy given the size of the catalogue and the iteration through many paths in it. In addition, daily backups are kept using slave servers that stay in sync with the master.

## 3. CVMFS pull-based solution through Apache Cassandra as main backend

The new scalable global grid catalogue consists in having Apache Cassandra [4] as main DB backend handling the ALICE Catalogue information, which means re-implementing the hierarchy on the NoSQL data model paradigm. In order to achieve this, the application workflow (query analysis) will be combined with the conceptual data model into the logical data schema, and implemented in the physical data model.

For end-users, a CVMFS interface will offer a browse-able catalogue with a familiar file system view of the contents. Any virtual organization could use such an interface as their own catalogue browser. On client demand, snapshots of the Cassandra database will be generated and distributed to these clients using the existing well-known and reliable CVMFS infrastructure, allowing for caching of the content along the way. Accessing the files requires interaction with the user credentials and the authentication/authorization services specific to each virtual organization, while keeping a common interface to access file metadata, which is one of the main goals of the project.

In addition, we study the utilization of new fast memory technologies i.e. Intel Optane, providing yet another type of cache layer to store objects in its DRAM form, and/or persistent storage in its SSD form. A schema of the new catalogue components can be seen in Figure 2.

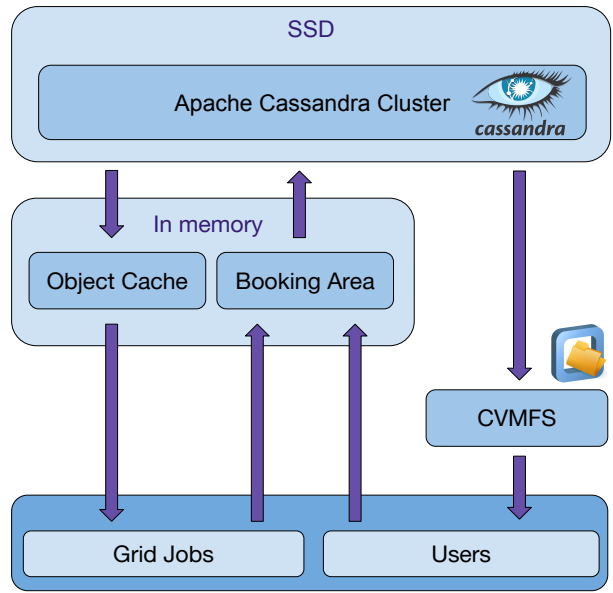

**Figure 2.** Global schema of the components

*3.1. Workflow explanation*

On one end, we have the entities that make use of the services, thus needing database information, in this case, grid jobs and users. Grid jobs have a set of input files that normally consist of macros, OCDB (conditions data) snapshots, other code files or in general any file that the submitter wants to add to the job. In addition, jobs that do analysis will access experiment data from the analysis framework. In our project schema, the metadata information needed as well as the authorization envelopes (keys to read/write from/to storage elements) will be prepared by the central services while processing the job insertions into the TaskQueue (grid jobs queue) and stored in a table, on a per waiting job basis, so that it can be queried quickly. Both the metadata and envelopes will be as well cacheable content, and will be sorted when the job starts, depending on the site it started on (for efficiency we usually run jobs where their data is). Sometimes extra files will be used by jobs, such as OCDB entries, the full mechanism that we currently support will still work, meaning that any job can ask for any file and be returned the list of locations optimally sorted related to its own location, it does not matter where the file is in the file hierarchy. Standard users fit in the use-case mentioned above. However, they will use CVMFS to have a UNIX-like view of the ALICE Catalogue. Snapshots of the hierarchies will come from upstream on demand, querying the Cassandra database backend. We generate SQLite content that will be sent to the squid server (HTTP request caching) next to the CVMFS client, and this client will retrieve the data from it, finally returning to the user the desired information. The process appears transparent to the end-user. CVMFS communication to squid and upstream uses the HTTP protocol, which helps avoid conflicts with firewalls.

When new files are to be added to the catalogue, first a booking of their logical file names is made. We will use the Booking Area for this, which must be non-volatile, since there has to be consistency with the database, in contrast to the Object Cache, that can go away without causing consistency issues. Once the initial booking is done, the client uploads the files to the corresponding storage element(s). To achieve this, the client needs to ask the central services for permission: check the user's file quota, find the best location based on storage element metrics such as space utilization, geographical position and network data, creation of envelopes with internal paths, MD5 checksums and more. When access is granted, the client uses the information provided by the central services, including the envelope, to trigger the remote copy and verifies that the file in the final destination is correct. An envelope is encrypted with the central services private key, while storage elements have the corresponding public key to decrypt the envelope and thus verify the legitimacy of the requested operation. Once the operation is finished, the final commit to the catalogue happens and the Booking Area reservation is released. In some cases, the final commit depends on certain events. For example, for jobs finishing with certain errors some files will be kept for a while, but by default not registered in the catalogue, unless the person in charge of the jobs needs them, normally for debugging. The Booking Area will be used for deletion of files from storage as well.

Lastly, at the other end, we have the central services themselves, in charge of providing all the information and operational support that the clients need. They use the database to perform multiple operations: catalogue maintenance, splitting jobs according to various strategies, calculate and refresh quotas, software packages management and more. The grid framework will query Cassandra directly, and in some cases interact with the cache. There are known use-cases where some values used by central processes will soon be used by jobs. Figure 3 below shows the read workflow for users and jobs.

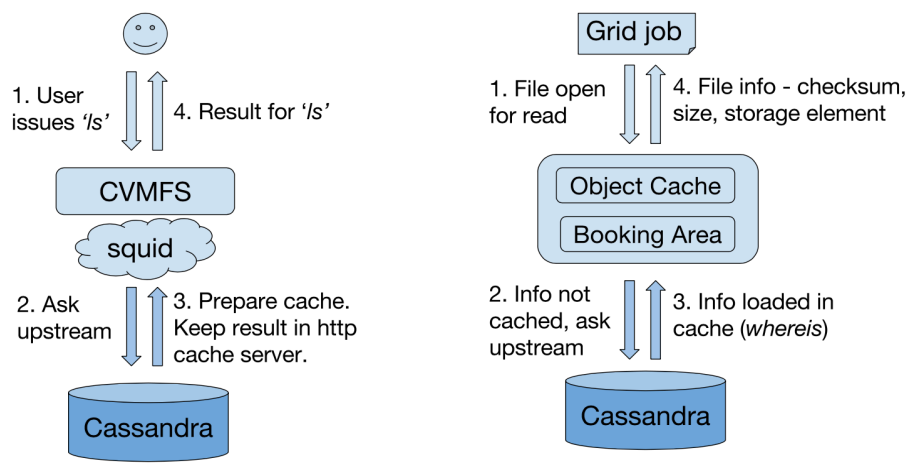

**Figure 3.** User (left) and job (right) read a file

## 4. Technology specifications

### 4.1. Apache Cassandra

Cassandra is a free and open-source distributed database for managing large amounts of structured data across many servers. It was first developed by Facebook, based on Google's BigTable and Amazon's Dynamo. It provides what other RDBMS or even NoSQL DBs can't: continuous availability, linear scalable performance, operational simplicity and easy data distribution across multiple data centers or clouds, while being a highly available service and having no single point of failure.

The architecture is designed as a ring, and replicates data across multiple nodes, following a masterless paradigm, and allowing failed nodes to be replaced with no downtime. This moves in a different direction than the traditional master-slave configuration or a manual and difficult-to-maintain sharded architecture.

From the technical point of view, the main challenge for the ALICE catalogue administrators will be to migrate from a RDBMS schema to the new Cassandra model, based on the combination of application workflow and conceptual data model. In this type of NoSQL system, the final implementation relies on the queries needed in the application. AliEn has a more or less constrained and well-known set of critical queries and the focus will be to optimize the DB model to have excellent performance in such queries, while keeping all the functionalities currently in the system. In addition to the performance, we have to keep it mind that at the same time, we need to have consistent information provided to the clients. Cassandra offers different tunable levels of consistency. We will always aim for strong consistency, and this is achieved by asking Cassandra to return the result agreed by a majority of nodes. This is especially important when we want to read a value that was just inserted or modified.

*4.2. Intel 3D Crosspoint (Optane)*
Intel Optane [5] is a great breakthrough in memory design. Its technology combines the revolutionary 3D XPoint non-volatile memory media with an advanced system memory controller. Intel's first benchmarks show a factor 8 data-rate-performance boost compared to their best NAND SSD on the market, while also having a much lower and stable latency (9 μs in their first tests) and 10 times more data density.

Intel announced they will provide SSDs based on this technology, as well as DRAM DIMMs, both persistent. We consider the utilization of this new memory in its SSD form for persistent information storage (DB backend, Cassandra), since the total expected volume is big. For the cache and booking area space we consider the DRAM version, more performant but costly, where we do not need to support such big data spaces. The cache will contain objects, possibly in a key-value fashion, loaded on demand, to filter many queries away from the database. The content of such a cache can be lost without causing problems to the workflows. The booking area will keep data that is temporarily needed by the Grid framework while the final commit has to wait for certain operations to happen, a kind of exchange area. It is relevant that in this case the memory is persistent, since it would allow the system to keep track of the files uploaded to Storage Elements at all times. Still, it would not be a major problem if sometimes the content were lost: affected jobs can just be rerun, while periodical consistency checks between the catalogue and the content of the Storage Elements will reveal mismatches to be rectified. Since the booking area is highly used by many concurrent tasks, having such a fast and common booking area will avoid synchronization problems within Cassandra and speed up the whole file registration process.

*4.3. CernVM-FS*
The CernVM-File System [6] provides a scalable, reliable and low-maintenance software distribution service. It is a well-known and widely used tool in the WLCG and users are familiar with it. It is implemented as a POSIX read-only filesystem in user space (a FUSE module). Files and directories are hosted on standard web servers (backed by squid servers) and CernVM-FS uses outgoing HTTP connections only, thereby avoiding most of the firewall issues of other network file systems.

The aim of this project is to use the established CVMFS infrastructure to present users with a browse-able AliEn File Catalogue with the familiar view of a standard filesystem. In order to achieve this, there are currently some developments to be made: authentication and authorization within the CVMFS client and/or by catching system calls and overwriting client commands; a DB to SQLite snapshot creation plugin to load Cassandra content into CVMFS; and a more fine-grained way to manipulate CVMFS cached data.

## 5. Initial benchmarking results

In order to evaluate the performance and possibilities offered by the Cassandra backend, we tailored a set of benchmarks to compare our current MySQL solution to our first iteration over Cassandra, and also to prove the linear scalability of the latter.

### 5.1. Database hardware and layout

The MySQL server used for the benchmarking is an HP ProLiant DL380 Gen9 with around 800GB of RAM, 20 physical cores and spinning disks in RAID6. From the database logical schema point of view, the file system hierarchy is split into many tables, both for the Logical File Names (LFN) and Physical File Names (PFN). In the first case, the separation comes from the path of the file. All files in the same table share a common initial part of the path. We have an index table to point to the right table to access depending on the LFN. This way of splitting allows us to isolate hierarchies so they don't interfere with each other, and to keep tables healthy size-wise. Regarding the PFNs, they are timestamp based and new entries are appended to the newest table. When the desired size limit is reached, a new table is created. Normally an LFN can be related to several PFNs, since in most cases we replicate the important data. The way LFNs and PFNs are linked is through a unique identifier (GUID).

In the case of the Apache Cassandra instance, we have created a 5-node ring with rather homogeneous servers. They are HP ProLiant DL380p Gen8 with 100 to 400 GB of RAM, 16 to 24 physical cores and spinning disks in RAID6, which means these are a very similar type of server compared to the MySQL master but with less RAM per node. As we mentioned before in this article, we need strong consistency in the data handled by the database, and Cassandra has tunable levels to achieve this. In our layout, the formula would be $WC + RC > RF$, where $WC$ is the write consistency level, $RC$ is the read consistency level and $RF$ is the replication factor. Since at the same time we want to keep our data safe, we need a high enough $RF$. Choosing a $RF$ of 3 would mean that we have 3 copies of each entry distributed in our cluster, and using Cassandra's *QUORUM* consistency both for reading and writing will give us the desired consistency. For very static areas of the namespace we could use ONE, which returns the first result from one of the nodes. In this case, 2 out of the 3 replicas have to agree on each read and each write operation, guaranteeing that we always get the latest results back. *LeveledCompactionStrategy* and compression via *LZ4* are used for optimizing reads.

The whole hierarchy is inserted into a single column family that combines LFNs and PFNs. We make use of an extra field to mimic the functionality of symbolic links, previously done by pointing from the PFN to the GUID of the parent LFN, thus having extra unnecessary entries in the database. Examples of the content of the database for different entry types can be seen in Table 1.

Table 1. Cassandra table schema and entries examples

| Field name | File | Symbolic link | Directory |
| --- | --- | --- | --- |
| **Path (PK)** | /alice/ | /alice/ | /alice/ |
| **Child (PK)** | file1 | link1 | dir1 |
| **Ctime** | 2016-11-15 15:46:00 | 2016-12-14 15:48:10 | 2016-11-11 10:10:05 |
| **Owner** | aliprod | aliprod | admin |
| **Gowner** | aliprod | aliprod | admin |
| **Jobid** | 0 | 0 | 0 |
| **Link** | | /alice/file1 | |
| **Md5** | ee31e45…a907ba51 | | |
| **Perm** | 755 | 755 | 755 |
| **Size** | 12152 | | |
| **Type** | f | l | d |
| **Pfns** | [{se1,pfn1},{se2,pfn2}] | [] | [] |

For the record a PFN example follows:

*root://castor.cern.ch:1094//10/33903/76cebd12-76a0-11e6-9717-0d38a10abeef*

Where *root://* is the protocol used in the storage, in this case Xrootd. It is followed by the address of the storage contact server, *castor.cern.ch:1094*, and finally we have the location within the storage file structure for the ALICE partition, *//10/33903/76cebd12-76a0-11e6-9717-0d38a10abeef*.

### 5.2. Benchmark specification

The Apache Cassandra cluster was populated with a subset of the original ALICE File Catalogue. In addition, a catalogue artificial generator was developed, creating mock entries with randomized values and following a sensible hierarchy. As reading and exporting the hierarchy from MySQL means reading the data from disk and sending it over, is a much slower operation than generating LFNs and PFNs artificially in CPU intensive tasks.

The insertion times were tested for both MySQL and Cassandra. We start from an empty table or column family and insert chunks of approximately 50 million entries. The idea was to track the evolution of the insert times depending on the total size of the namespace.

In a similar fashion and in parallel with the insertions, we benchmarked the reading times by implementing the `whereis` command from AliEn for the Cassandra model. This operation retrieves all the logical and physical metadata about each file. In other words, both the LFNs and PFNs.

In order to send requests, 20 intensive simultaneous clients query the databases at a frequency of several hundred Hz.

The last benchmark ran only on the Cassandra cluster and measured the evolution of the operations per second (throughput) depending on the number of clients issuing requests, to show the linear scalability of the database.

### 5.3. Benchmark results

Apart from what was commented on previous points, an additional remark is that we have decided to run queries in Cassandra with both strong and weak consistency to verify if the difference is very significant.

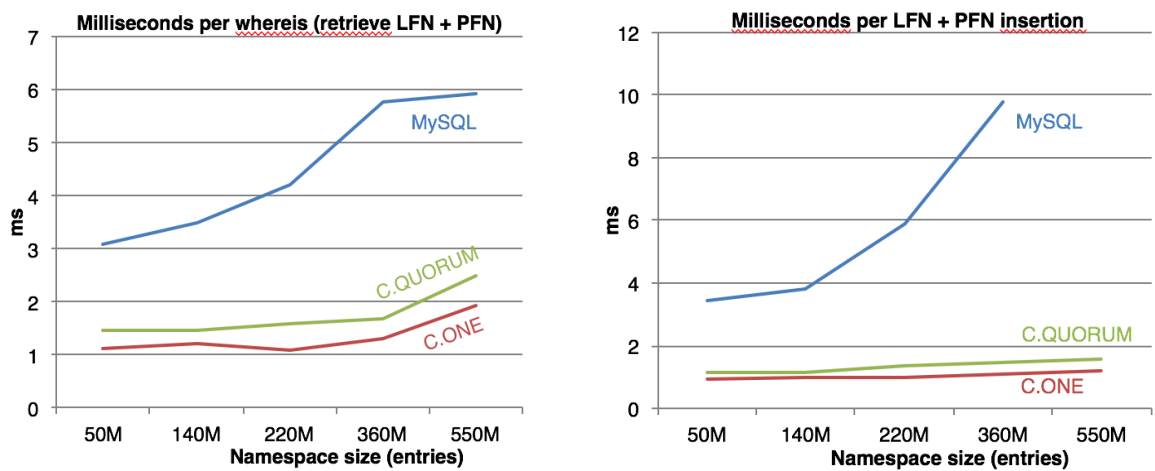

**Figure 4.** Results of retrieving (left) and inserting (right) logical and physical metadata of ALICE files

Figure 4 shows that, as expected, insertions in Cassandra are very cheap. This is due to its write path, where all the writes are kept in memory, and this structure kept in the RAM is flushed to disk when a size threshold is reached, avoiding all contact to disk meanwhile.

Regarding the read benchmark, results are very positive as well for this use case, where we can point to a specific entry by using the partition key/s in the *whereis* part of the query, which is very efficient in Cassandra.

For both tests, namespace size is not a problem in Cassandra.

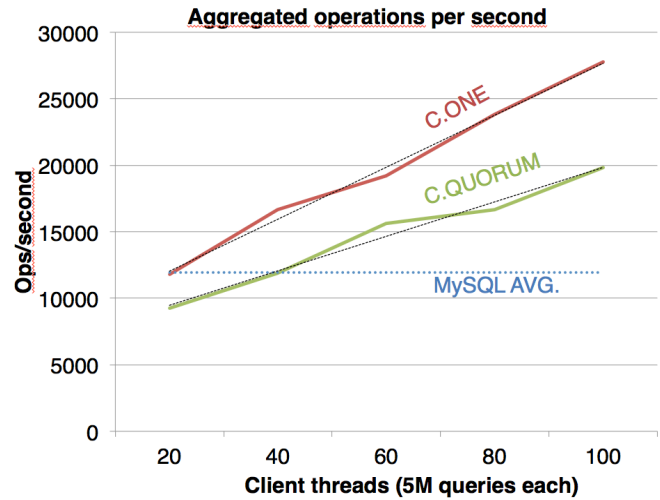

**Figure 5.** Throughput depending on the number of clients.

Linearity on operations per second is clearly shown in Figure 5, with a fluctuation of 20 to 100 threads. As mentioned before in this document, we can easily add nodes to our cluster, thus distributing queries among more servers. This feature allows the database keep up with the load in case we reach the limit of operations per second that the cluster can handle.

## 6. Conclusions and future work

Apache Cassandra is a good alternative for the ALICE File Catalogue, covering all the key points that the current MySQL cannot fulfill while providing all the required functionalities. The next challenge will be to port the rest of the query workflow still not implemented into the NoSQL paradigm. Heavy benchmarking and operations control will be a very important part of the work given our current and future requirements. Together with the database backend migration, the grid framework will evolve to adapt to such new software and optimize its internal workflows to keep up with the load. Study of arising hardware technologies might facilitate some of the steps described in this paragraph and make significant changes in the final deployment, even if our system is thought out to be scalable by design.

**References**
[1] ALICE Collaboration – A Large Ion Collider Experiment. http://alice-collaboration.web.cern.ch
[2] Worldwide LHC Computing Grid – http://wlcg.web.cern.ch
[3] ALICE Environment – AliEn – http://alien.web.cern.ch
[4] Apache Cassandra distributed database – http://cassandra.apache.org/
[5] Intel Optane, memory breakthrough – http://www.intel.com/content/www/us/en/architecture-and-technology/intel-optane-technology.html
[6] CernVM Filesystem - https://cernvm.cern.ch/portal/filesystem